\documentclass{elsart5p}
\usepackage{graphicx}
\usepackage{amssymb}

\begin{document}

\begin{frontmatter}

\title{Galactic Sources of High-Energy Neutrinos: Highlights} 
\author{Francesco Vissani${}^1$ and Felix Aharonian${}^2$}\thanks{We thank ML.Costantini, N.Sahakyan and F.Villante for  collaboration, P.Blasi and P.Lipari for pleasant and important discussions, 
T.Schwetz for an explanation on $\theta_{13}$ and F.Halzen for 
an interesting public discussion  on Fig.~\ref{fig8} at NUSKY meeting (ICTP).}
\address{${}^1$INFN, Gran Sasso Theory Group, Assergi (AQ) Italy}
\address{${}^2$Dublin Institute for Advanced Studies, 
31 Fitzwilliam Place, Dublin 2, Ireland}

\begin{abstract}
We overview high-energy neutrinos from galactic sources, transparent to their  
gamma-ray emission. We focus on young supernova remnants and in particular on RX J1713.7-3946, discussing expectations and upper bounds. We also consider the possibility to detect neutrinos from other strong galactic gamma-ray sources as Vela Junior, the Cygnus Region and the recently discovered Fermi Bubbles. We quantify the impact  
of the recent hint for a large value of $\theta_{13}$
on  high-energy neutrino oscillations.
%
\end{abstract}

\end{frontmatter}






\section{Context, motivations and assumptions}
\begin{figure}[b]
\centerline{\includegraphics[width=7.1cm,height=4.cm,angle=0]{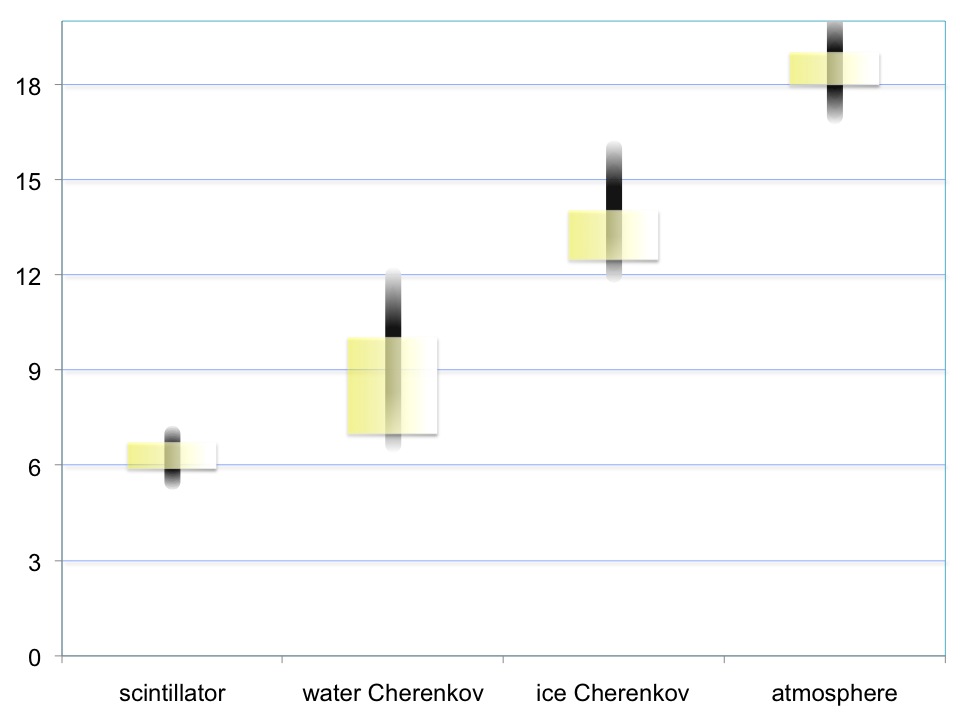}}
\caption{\em\footnotesize  Energy ranges of various types of 
neutrino telescopes, in $\log_{10}(E/eV)$; note that they
span more than 10 decades in energy. In this paper, we are concerned
with the third item, that after IceCUBE we identify  
with the technology `ice Cherenkov'.
\label{fig0}}
\end{figure}

The successes of low energy neutrino astronomy and the discovery of neutrino oscillations added momentum to the search for high energy neutrinos from cosmic sources, initiated long ago with the
theoretical proposals of Zheleznykh, Markov and Greisen \cite{markov}; see Fig.~\ref{fig0}. 
The first km$^3$-class detector, IceCUBE and the smaller ANTARES (following MACRO, BAIKAL, AMANDA) have not yet revealed the first signal, yet the excitement remains high. However, nowadays it becomes clear that the search for high energy neutrino sources is difficult. In this paper, we attempt to examine certain generic and specific expectations for such a search, in the hope to contribute to the scientific planning of the future experimental activities. We are interested in the possibility to proceeding further, by building a new telescope of km$^3$-class in the Northern Hemisphere--as Km3NET in the Mediterranean Sea or GVD in Lake Baikal.

The fact that the first explorations of the high-energy neutrino sky, that did not reveal any signal, contradicted  the over-optimistic expectations may lead one to doubt that we can reach
reliable predictions. But, even if some degree of diffidence toward theory is healthy, we are convinced that the process of formulating expectations is an important step toward understanding. 
We would like to clarify our view by proposing some issues: 
{\em Do we really need to have high expectations for high energy neutrinos?} Perhaps not, although one should keep in mind that the surprises often happen in astronomy. 
As recent example in this regard are the discovery of the Fermi Bubbles and the variability of the Crab Nebula.
{\em Are expectations useful?} Of course, yes; good predictions are precious for experiments, but also 
reasonable expectations eventually contradicted are not useless.
{\em Do we have some relevant and succesful precedent?} 
Yes, many: 
High-energy $\gamma$-ray sources have been predicted in the fifties (Morrison).
For solar neutrinos, we had predictions with errorbars since the sixties (Bahcall).
For supernova neutrinos, quantitative expectations were available even before SN1987A
(Nadyozhin).

\begin{figure}[t]
\centerline{\includegraphics[width=4.2cm,angle=0]{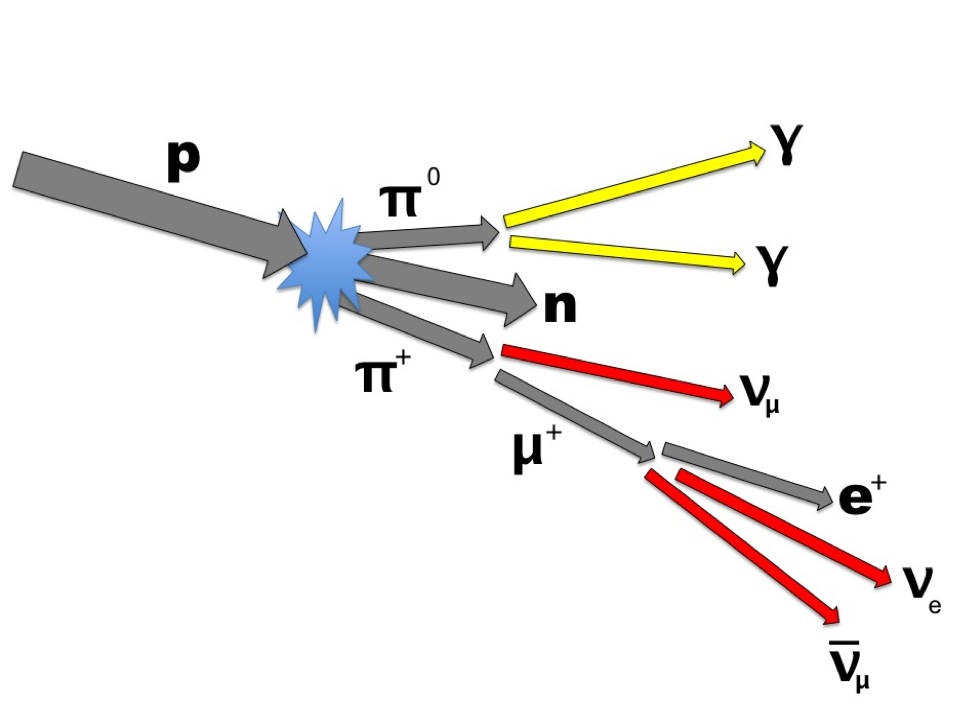}}
\caption{\em\footnotesize  \label{fig1}
We can exploit the strict connection of secondary muon neutrinos and $\gamma$-rays  produced by collisions of cosmic rays above 10 TeV, assuming that the $\gamma$-rays are not absorbed or strongly modified.} 
\end{figure}

We adopt a rather common astrophysical attitude: the observation of high-energy neutrinos is very demanding but perhaps possible and would amount to an unambiguous signal of cosmic ray 
collisions--hopefully, those in their source. We simplify our task by restricting our attention on the specific but important class of sources, those transparent to their gamma rays.
We will show how to proceed
towards precise expectations and eventually observations of neutrinos,
limiting the use of theory inputs and just by the help of  $\gamma$-ray observations.
This aspect is important enough to be emphasized, as we do with Fig.~\ref{fig1}.
In the next section, we quantify the connection between gamma and neutrinos. 

\section{Expected neutrino intensity and  signal}
If we measure the very high $\gamma$-ray emission from a cosmic source, and if we attribute it  
to cosmic ray colliding with other hadrons, it is straightforward to derive the muon neutrino flux.
In fact, when both the neutrinos and {\em unmodified, hadronic} $\gamma$-rays are linear 
functions of the cosmic ray intensity, they are linked by a linear relation  \cite{fv08}: 
\begin{eqnarray}
I_{\nu_\mu}(E)&=&0.380\ I_\gamma\left(\frac{E}{1-r_\pi}\right)
+0.013\ I_\gamma\left(\frac{E}{1-r_K}\right)\\[-2ex] \nonumber
&+&\int_0^1 \frac{dx}{x} {K_\mu(x)}
I_\gamma\left(\frac{E}{x}\right)
\end{eqnarray}
where the coefficients are determined by the  
amount of mesons produced and by their decay 
kinematics.
The first and second contributions are due to direct decays into neutrinos 
and the third one to $\mu$ decay, the kernel $K_\mu(x)$ being
\begin{equation}
\left\{
\begin{array}{ll}
x^2(15.34-28.93 x) & 0<x<r_K \\[-1ex]
0.0165+0.1193 x+3.747 x^2-3.981 x^3 & r_K<x<r_\pi \\[-1ex]
(1-x)^2 (-0.6698+6.588 x) & r_\pi<x<1 
\end{array}
\right.
\end{equation}
where $r_x=(m_\mu/m_x)^2$ with $x=\pi,K$.
Similar formulas describe antineutrino production; 
the effects of the oscillations are included \cite{fv08}, compare also with
\cite{fv06} and see the Appendix for an updated discussion. 


\begin{figure}[t]
\centerline{\includegraphics[width=5.4cm,angle=0]{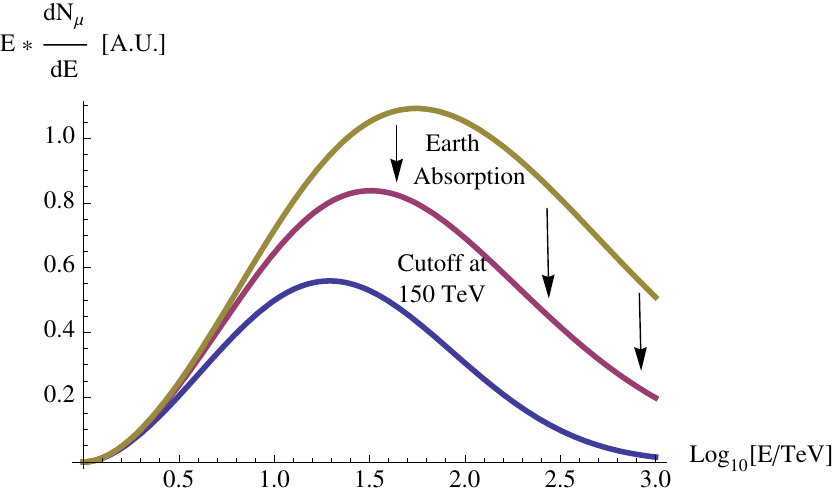}}
\caption{\em\footnotesize 
Distribution of $\nu_\mu$ leading to muons, 
assuming 
$E^{-2}$ primary spectrum (upper curve); then, including Earth absorption, 
for a source at $\delta=-39^\circ$ as seen from ANTARES (middle curve); then 
with a spectrum $E^{-2} \exp(\ {-\sqrt{E/150\mbox{\rm\tiny\ TeV}}}\ )$
(lower curve), i.e., with primaries cutoffed at $\sim$3 PeV. \label{fig2}}
\end{figure}

At this point, calculating the muon signal is just applying a 
{\it standard} procedure. First, one evaluates the probability of 
converting muon neutrinos into muons
\begin{eqnarray}
P_{\nu_\mu\to \mu}&=&\int_{E_{th}}^{E}  
dE_\mu \frac{d\sigma_{cc}}{dE_\mu} R_\mu/m_n \\[-1ex]
&&\mbox{[say, $10^{-35}$ cm$^2\times /m_n \beta\sim 10^{-6}$]} \nonumber
\end{eqnarray}
Unless stated otherwise, we will use $E_{th}=1$ TeV in order to cope with 
atmospheric neutrino background; see \cite{tevo} and Fig.~\ref{fig0}.
Then, one should calculate the effective neutrino detector area, proportional 
to the previous probability,
to the  physical area of the muon detector (that in general depends on the zenith angle $\theta$), 
 and to the probability that the Earth 
absorbs neutrinos: 
\begin{eqnarray}
A_{\nu_\mu}&=&A_\mu(\theta)\times P_{\nu_\mu\to \mu}(E,\theta)\times e^{-\sigma\ z /m_n}\\
\nonumber &&
\mbox{[say, 1 km$^2\times 10^{-6}\sim 1$ m$^2$]}
\end{eqnarray}
When $E\sim 10$ TeV, $s\sim 2 m_n E\sim Q^2> M_W^2$, 
thus this is the energy where the cross section begins to grow more slowly. 
The Earth
absorption begins at 
$E\sim\mbox{few}\cdot 100\mbox{ TeV}$, when 
$\sigma(E)\sim m_n/(R_\oplus\bar\rho_\oplus)\sim 5\cdot 10^{-34}$ cm$^2$, see 
Fig.~\ref{fig2}.

\section{Neutrinos and gamma rays}
\begin{figure}[t]
\centerline{\includegraphics[width=6cm,angle=0]{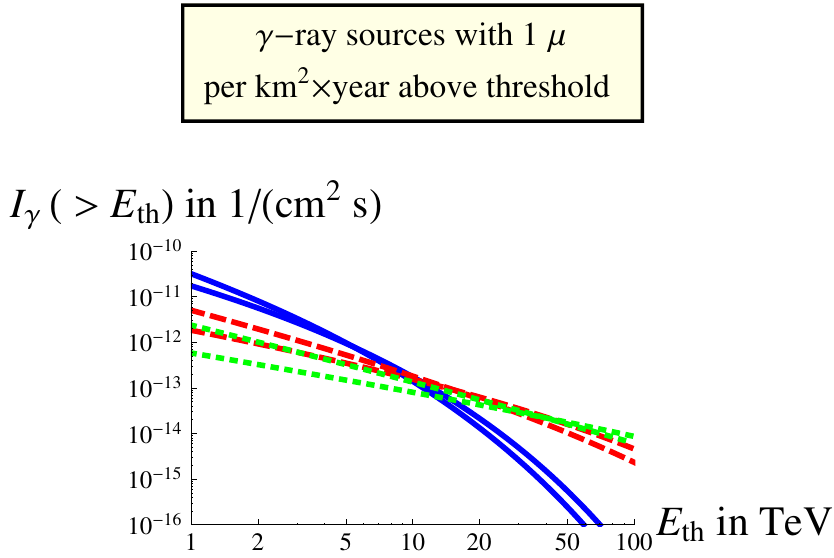}}
\caption{\em\footnotesize 
$\gamma$-ray integral intensities, corresponding to a signal of 
 1 muon/km$^2$yr and above a threshold $E_{th}$, evaluated assuming 
 that the sources are {transparent} to their gamma rays. See the text, and 
  compare with \cite{narek}, where we show the differential 
 intensities instead. \label{fig3} }
\end{figure}

If one assumes a power-law distribution with an exponential cutoff
 for the primary cosmic rays, then the spectrum of $\gamma$-rays has the same power-law index but drops slower \cite{kelner}
\begin{equation}
I_\gamma \propto E_\gamma^{-{\alpha}} 
\cdot \label{mpl}
\exp(-\sqrt{E_\gamma/{E_{c}}}).
\end{equation} 
Typically, in $\gamma$-ray sources 
$\alpha$ varies between $1.8-2.2$ and  and $E_c$ varies from 1 TeV to 100 TeV.
Assuming $\gamma$-ray transparency we can calculate the critical (minimal) intensity 
that corresponds 1 muon/km$^2$ year above 1 TeV; this is given in Fig.~\ref{fig3},
where we consider $\alpha=1.8,2.2$ and  $E_c=1,30,1000$ TeV.
Interestingly, the integrated flux above 10 TeV is almost the same in all these cases
\begin{equation}
I_\gamma(>10\mbox{ \rm TeV} )=(1- 2)\times 10^{-13}/\mbox{\rm(cm$^2$ s)}
\end{equation}
Note that the present Cherenkov telescopes are less sensitive above 10 TeV,
due to the small collection area. Thus we ask how to detect these fluxes. In order to
collect $\ge 100\ \gamma$'s in a reasonable time,  we need a km$^2$ area:
\begin{equation}
\mbox{Exposure}=L^2\times T\sim 2\times \mbox{ \rm km}^2\times 10\mbox{ \rm h}
\end{equation}
e.g., a 10$\times$10 Cherenkov telescopes array, or one
dedicated EAS array.
A large area $\gamma$ apparatus, such as
CTA or even a custom instrument, could be then worthwhile even 
for neutrino community, and should cost much less than a neutrino-telescope.

\begin{figure}[b]
\centerline{\includegraphics[width=4.cm,angle=0]{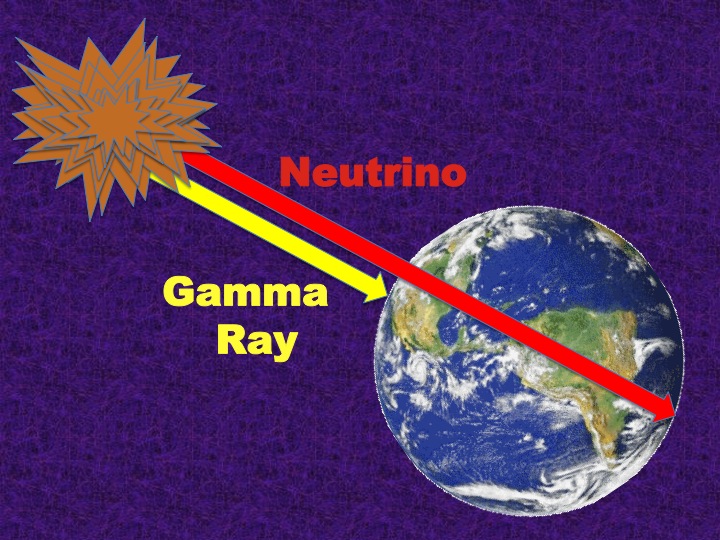}}
\caption{\em\footnotesize 
The main particularity of neutrino-telescopes is that they
look downward; in fact, due to atmospheric $\mu$ background,  
$\nu_\mu$ from cosmic sources
are preferentially detected from below.\label{fig4}}
\end{figure}

Now that we have compared  the energy-response of 
gamma- and neutrino-telescopes, we would like to 
emphasize the fact that 
$\gamma$ and $\nu_\mu$ views are 
complementary, as suggested by Fig.~\ref{fig4}.
The maximal complementarity would be obtained with 
two detectors in antipodal locations.
Stated in more quantitative terms, a steady source 
at declination $\delta$ is seen from a detector at latitude $\phi$ for a fraction of time:
\begin{equation}
f_\gamma ={\mbox{Re}[\cos^{-1} (-\tan\!\delta\tan\!\phi)]}/{\pi};
\end{equation}
the fraction of time for neutrinos is just $f_{\nu_\mu}=1-f_\gamma$.
A concrete way to illustrate this feature is to note that 
a hypothetical $\nu_\mu$ (resp., $\gamma$) 
emission from Galactic Center, $\delta\approx -30^\circ$, 
is visible from North (resp., South) Pole.

\begin{figure}[t]
\centerline{
\includegraphics[width=5cm,angle=0]{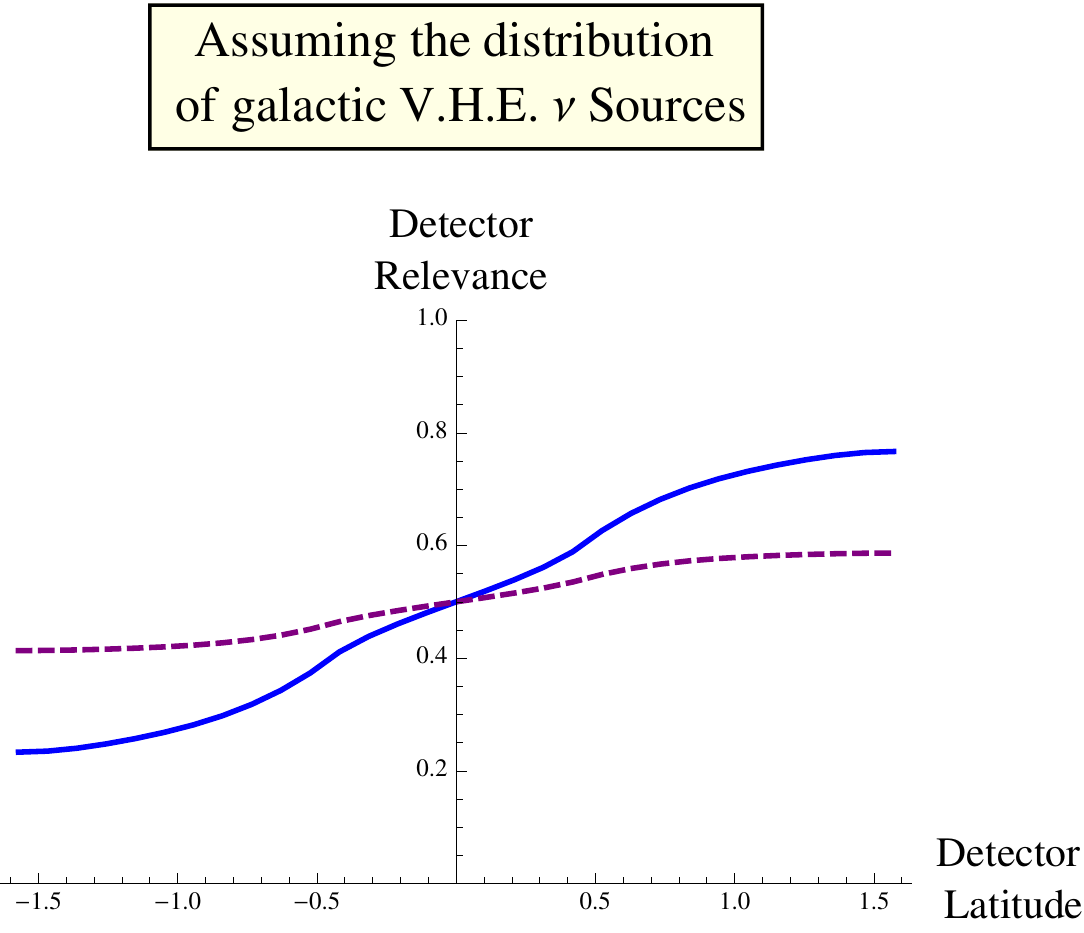}}
\vskip-3mm
\caption{\footnotesize\em Continuous line: detector relevance, calculated assuming the 
galactic mass distribution. Dashed: same,
weighted with $1/r^2$ to account for the fact that nearby sources are easier to detect.
From  \cite{narek}.\label{fig6}}
\end{figure}

This argument can be generalized.
Suppose that the galactic sources of TeV neutrinos trace the mass distribution, and 
calculate the relevance of a detector as a function of its latitude.
The matter of the Galaxy is mostly located in the region $\delta<0$, i.e., 
below the celestial equator. 
Thus, a telescope at the latitude of NEMO has a priori 
2.9 better chances (or 1.4, weighting 
the mass distribution with $1/r^2$) 
to see galactic neutrino sources than IceCUBE as 
shown in Fig.~\ref{fig6} and argued in \cite{narek}.
These remarks show why we think that a 
large, high-energy neutrino 
detectors in the Northern Hemisphere, not only would 
complement the IceCUBE field of view, but would also  
be important to probe the existence of 
hypothetical galactic neutrino sources.

However these considerations, as general as they are, leave 
something to be desired. For instance,  the
use the average galactic 
matter distribution to infer the source neutrino distribution is doubtful, also 
because HESS sky-maps show only few intense $\gamma$-ray sources.
In this regard 
it is more appropriate to study individual sources. 

In order to proceed, it is useful to formulate explicitly 
some pending question and doubt: 
(i)~Is $\gamma$-transparency a reliable hypothesis?
(ii)~Are we sure of the `point-source' hypothesis?
Similar as asking whether $\ll\!1^\circ$ pointing is really important
for very-high-energy $\gamma$ and/or $\nu$  telescopes?
(iii)~$\gamma$ above 10 TeV can help $\nu$ astronomy, and this would be 
a natural 	direction of progress anyway, but do we have any {guaranteed} aim for such a search? 
(iv)~Eventually, the true question is: How to separate
{leptonic from hadronic} gamma's? 
Neutrino identify CR collisions, but is this the
only way to proceed?

All these questions suggest to define more precisely 
which (type of) source we want to investigate.

\section{
Supernova remnants+molecular clouds}

The hypothesis that supernova remnants (SNR) are main 
sites of the acceleration of the galactic cosmic rays (CR) can and should to be 
tested.\footnote{Let us recall the main steps that consolidated this hypothesis:
1) Baade \& Zwicky's propose that 
supernova explosions and CR are connected;
2)  Fermi argues that the kinetic
energy of gas can convert into CR acceleration via magneto-hydrodynamical
processes;
3) Ginzburg \& Syrovatskii note that the kinetic energy injected by SNR
is 1-2 orders of magnitudes greater than the galactic energy losses
of CR;
4) several authors note that the shock wave 
present in astrophysical environments as SNR offers 
the conditions for efficient CR acceleration;
5) Bell \& Lucek et al.\ remark that the CR lead to an amplification of the relevant 
magnetic fields in the region of the shock wave, as observed;
6) the present efforts are directed to develop a full non-linear theory of 
CR and SNR possibly including MC.}
As indicated in ref.~\cite{d},  
the associations between SNR and molecular clouds (MC) provide the optimal conditions to produce intense neutrino- and $\gamma$-ray fluxes, that  thereby  can be used  as a tool to test the SNR paradigm.

The best hope to detect `hadronic' neutrinos is connected then to 
young SNR, i.e., of age less than 2000 yr. 
In fact, these are expected to contain the highest energy cosmic rays, and can thus produce energetic neutrinos that stand over the background of atmospheric neutrinos. 
In other words, we are interested in special types of 
SNR's as potential neutrino sources: the closest and the youngest, 
among those associated to molecular clouds. It should however be observed that core collapse supernovae, being connected to short living stars, are pretty naturally associated with sites of intense 
stars formation--i.e., molecular clouds.

Before analyzing a specific and important case of such a young SNR,
let us discuss in some generality the pros \& cons of the SNR+molecular clouds paradigm, outlined above.
Some support comes from the detected GeV $\gamma$'s from relatively old SNR, 
 as W28 and W44.  Moreover, gamma transparency usually holds. 
One can expect, a priori, that the closest  SNR should be at about 1 kpc; in fact, 
we have 1 new SN each many tens or years and the
size of the Galaxy is several tens of kpc.
Also, it is true that the CR acceleration above 
100 TeV is still an open theoretical problem but, on the other hand, 
it is a problem that can be approached observationally measuring gammas above 10 TeV. 

In any case, we should expect that in order to describe any individual SNR we 
need theoretical modeling anyway (and the presence of MC will not make the matter simpler)
plus as many multiwavelength observations as possible.

\begin{table}[t]
\begin{center}
\begin{tabular}{|c|c|c|c|}
\hline
Muon & Expected  & 1 sigma sta- & Atmospheric \\[-1.5ex] 
threshold & signal & tistical error & background \\ \hline
50 GeV & 5.7 & 6\% & 21 \\
200 GeV & 4.7 & 7\%  & 7 \\
\bf 1 TeV & \bf 2.4 & \bf 10\% & \bf 1 \\
5 TeV & 0.6 & 30\% & 0.1 \\
20 TeV & 0.1 & 100\%  & 0.0\\ \hline
\end{tabular}
\end{center}
\caption{\em\small Dependence on the threshold of the  
number of signal muons per km$^2$ per year 
from RX J1713.7-3946 and assuming the hadronic hypothesis. Also quoted the estimated error from HESS statistics and the estimated background. From \cite{fv08}.\label{tab1}}
\end{table}

\section{The SNR RX J1713.7-3946: a case study}

In this section, we will discuss the object RX J1713.7-3946, in particular, in connection with the possibility that this is an intense source of neutrinos above TeV.  The nature of the object, its age and its distance are still debated issues; however, the identification of the location of this source of non-thermal X ray emission \cite{asch} with the position of a historical supernova \cite{wang}
supports the idea that this is a SNR about 1600 years old. The distance is estimated around 1 kpc \cite{nanten}; thus, it is pretty close. Moreover, the 
observations indicate an existence of dense condensations inside the shell of RX J1713.7-3946 \cite{nanten}.
The most prominent one has $\sim 400$ $M_\odot$ and is 1/10 of the SNR angular size. 
Thus, it should have a volume of (1 pc)$^3$ and therefore a
column density of $20$  $\mu$m of lead, that cannot produce a significant attentuation of the gamma rays. It has been argued that the structure of this system of molecular clouds and the SNR shock waves are intimately connected \cite{fukui}.

Another feature of this SNR is that $\gamma$-rays up to 30 TeV or perhaps even more have been measured by HESS \cite{hess}. This is a pretty unique circumstance, in view of the limited area of the existing $\gamma$-ray telescopes that explored the TeV region but barely begun the exploration of the sky above few 10 TeV.  The spectrum is non-trivial:  it is 
not described by a simple power law, but rather by a  broken-power-law or by a modified-exponential-cut as in Eq.~\ref{mpl} and with  \cite{vv07}
\begin{equation} \label{sloo}
\gamma=1.79\pm 0.06\mbox{ and }E_c=3.7\pm 1\mbox{ TeV}
\end{equation}
this does not seem to contradict the expectations, but the cutoff
suggests that the cosmic rays with energies above 
150 TeV have already abandoned this SNR.

\subsection{Expectations within the hadronic hypothesis}
Thanks to HESS measurements, the 
upper bound on neutrino signal is now quite precise. 
The most detailed recent 
calculation \cite{fv08} 
deduced 
the $\nu_\mu$ and $\bar\nu_\mu$ from these data,
finding for a threshold of 1 TeV an induced muon flux of   
\begin{equation}
I_{\mu+\bar\mu}=2.4\pm 0.3 \pm 0.5 /\mbox{km$^2$ yr}
\end{equation}
In other words, the HESS data, with the hadronic
hypothesis, permit us a reliable evaluation of the expected fluxes--or more plausibly, 
the upper bounds--on high energy neutrinos. It is not possible to lower much the detection 
threshold since the background due to atmospheric muons increases rapidly \cite{tevo}: 
in fact, the flux of muon neutrinos decreases as $\sim E^{-3.7}$ in the relevant region,
as we see from Tab.~\ref{tab1}.

\begin{figure}[t]
\centerline{\includegraphics[height=6.5cm,width=3.4cm,angle=270]{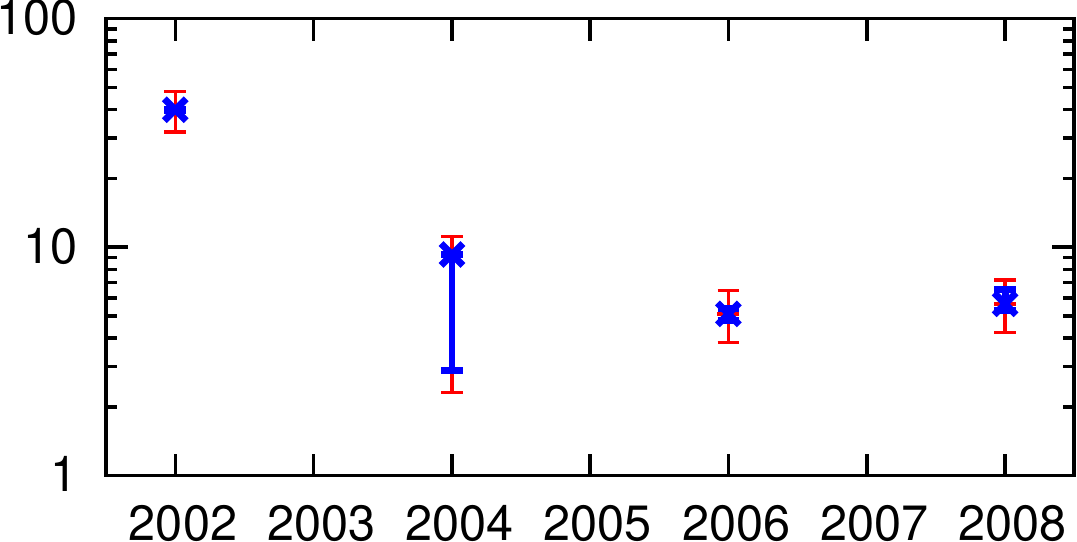}}
\caption{\em\footnotesize 
Various predictions for the muon flux per km$^2\times$yr expected from RX J1713.7-3946
and above 50 GeV. 
In blue, we quote the error deduced from the publications quoted in the text,  
in red, we quote 20\% systematic error. 
In the horizontal axis,  the year of the publication.
The reasons of the changes are discussed in the text.
From \cite{nuovoCim}. \label{fig8}}
\end{figure}


The expectations changed in the course of the time  
as shown in Fig.~\ref{fig8}; let us recall the reasons.
Ref.~\cite{halzen}:
The first work on high energy 
neutrinos from RX J1713.7-3946  used 
a bold extrapolation of  Cangaroo data,  and adopted a
simplified approach, subsequently improved.
Ref.~\cite{fv04}: the inclusion of oscillations,
Earth absorption, and livetime 
caused a factor of 4 decrease; moreover, the hypothesis of an early cutoff in the spectrum was 
remarked, estimating the maximum decrease in a factor of~3.
Ref.~\cite{fv06}: the results of HESS, which first 
revealed a cutoff in the spectrum, permitted us to quantify the 
decrease in about a factor of 2. 
Ref.~\cite{fv08}: None of the subsequent theoretical and observational improvements considered
introduced large changes, and the prediction became stable.

Two remarks are in order:
(i)~the last three predictions (or better, it is good to repeat it, upper bounds) agree within errors;
(ii)~for a fair comparison, the threshold has been set to 
the common value of $E_{th}=50$ GeV, but this value  
it is too low to permit background rejection.
As shown in Tab.~\ref{tab1}, this implies that the expected signal should be more than halved. 

\subsection{Fermi-LAT view at {\it GeV} and above}
However, the true point is whether the hadronic hypothesis holds true 
(i.e., whether the observed $\gamma$-rays come predominantly from CR collisions); 
crucial tests of this hypothesis (that are of course relevant to the matter of predicting the high energy neutrino signal) include the observations of $\gamma$-rays 
in various regions of energy. 

Thus it is very interesting that, recently, Fermi-LAT released their first results on RX J1713.7-3946 \cite{fermi}, that revealed a rather complicated 
environment:  
1)~there is a wide source  approximatively in the SNR 
location with spectrum 
\begin{equation}
I_\gamma \propto E^{-1.5\pm 0.1}_\gamma
\end{equation}
that results from measurements above 5 GeV and from 
an upper bound on the spectrum at 0.5-5 GeV; 
2)~there are several  point sources in the surrounding, including one
sloping as $\approx E^{-2.45}_{\gamma}$, outshining the wide source at GeV energies;
3)~finally, there is the diffuse background from the Milky Way.

\begin{figure}
\centerline{\includegraphics[width=6cm]{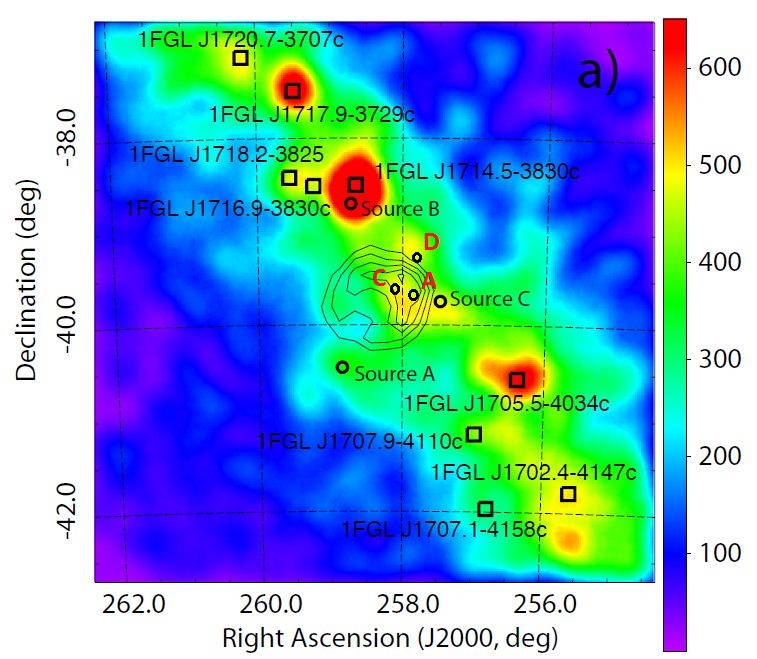}}
\caption{\em\footnotesize 
The region around RX J1713.7-3946 as seen by Fermi-LAT, superimposed with the location of the dense condensations named A,C,D seen by 
NANTEN--small circles marked by the red letters. 
The ``Sources A,B,C''--larger circles marked by black letters--are the new point 
sources found by Fermi-LAT instead. From \cite{fermi}.\label{figr}}
\end{figure}

On the one hand, the location of the extended source does not disagree with the location of 
the dense condensations 
as shown in Fig.~\ref{figr}; however, what seems most crucial for the 
interpretation is the slope of the spectrum, that is 
different from the one given in Eq.~\ref{sloo} and also harder than theoretically expected one,
assuming that the observed emission is due to CR collisions. 

Let us pause here to
recollect. 
From a purely observational point of view, 
it seems very important to understand better the emission below 5 GeV, and one should note that 
a spectrum like $E^{-1.7}_\gamma$ is not yet excluded firmly--which could be reconciled more easily, perhaps, with hadronic emission and a
{\em very} efficient acceleration. From a theoretical point of view, the spectrum could be leptonic   
($E_e^{-\gamma}\Rightarrow$ $E_\gamma^{-(\gamma+1)/2}$) 
or hadronic, with an energy-dependent penetration factor
($E_p^{-\gamma}\Rightarrow$ $E_\gamma^{-\gamma+1/2}$),
if primaries--electrons or protons, respectively--have $\gamma\approx 2$ \cite{fukui}.
Moreover, the lack of thermal X-rays could imply that the emission is leptonic if 
the medium is uniform, or that non-uniformity plays a 
major role \cite{ellison,fukui}.

Since we expect progress from the GeV region in the close future, we believe  
that the right attitude is: Wait and see.

\subsection{Next steps?}

\begin{figure}[t]
\centerline{\includegraphics[width=6cm,angle=0]{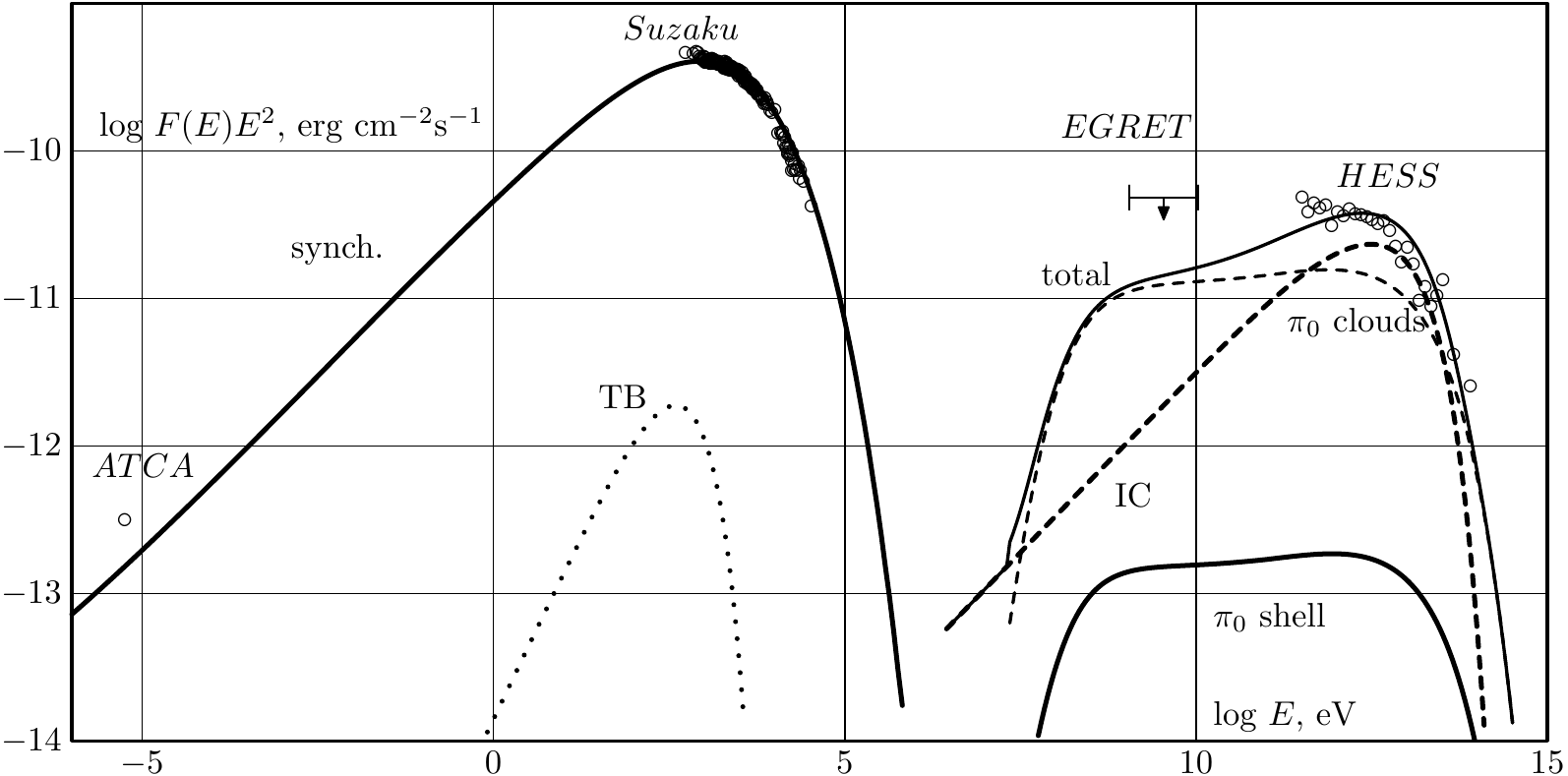}}
\caption{\em\footnotesize  Mixed composition of the spectrum of RX J1713.7-3946 as predicted by
the Zirakashvili \& Aharonian model. Figure from \cite{za}.\label{sss}}
\end{figure}

The possibility that the spectrum of RX J1713.7-3946 is composite was
discussed in a concrete theoretical model \cite{za}, see Fig.~\ref{sss}
(note however that the need to disentangle hadronic and leptonic emission was pointed 
out immediately, see e.g., the statement of Aschenbach in \cite{vulcano}).
This shifts the discussion on  
the need to {quantify} hadronic and leptonic $\gamma$ emission, rather than excluding one
model in favor of the other one. 
Consider the possibility that the spectrum is composite.  
If we want to check thoroughly the SNR+MC paradigm, 
thus improving the expectations on high-energy neutrinos, 
we should not study only the spectrum: we should also measure precisely the arrival directions 
of the high-energy $\gamma$'s, testing correlations with molecular clouds. This is especially important 
for $\gamma$'s at 10 TeV and above, that are unlikely to be leptonic. 
Is it possible to use the existing data for this purpose? HESS has 
about 500 events above 30 TeV, but 70\% of them is expected to be 
cosmic ray background (i.e., noise) \cite{hess}.
Then, in order to perform this check, we should wait for CTA or for an equivalent 
km$^2$ class $\gamma$-ray telescope, sensitive above 10 TeV, 
and with good angular pointing.

\section{Other possible sources of high energy neutrinos}

\noindent  {\bf Vela Junior:}
Above the bound necessary to have more than 1 $\mu$/(km$^2\times$yr) \cite{narek}, namely,
\begin{equation}
I_\gamma(20\mbox{ TeV}) = (2 - 6) \times 10^{-15} /\mbox{TeV cm$^2$ s}
\end{equation}
there are 2 more young SNR, Vela Jr and Vela X, observed by HESS.
The first one, also known 
as RX J0852-4622, is a young shell-type SNR, just as RX J1713.7-3946, with
angular size $2^\circ$. The specific parameters are however quite unknown; the most 
common values of the 
estimated age are of 660-1400 yr, implying a distance of 0.26-0.50 kpc.
Actually, this source is even more intense than RX J1713.7-3946 in $\gamma$-rays:
\begin{equation}
I_\gamma(20\mbox{ TeV}) = (1 - 3) \times 10^{-14} /\mbox{TeV cm$^2$ s}
\end{equation}
but 20 TeV is the last point presently measured by HESS.          
HESS spectrum $\propto E^{-2.1},$ whereas Fermi-LAT's preliminary measurement
gives $\propto E^{-1.9}$ \cite{razz}, which could indicate a spectral break.

\noindent  {\bf Cygnus Region:}
At 1.7 kpc from us, there is a large star forming region of $100,000 M_\odot$ mass in the direction of  
Cygnus constellation. This is remarkable target for cosmic ray collisions, even if we do not know for sure which should be the associated source of cosmic rays.
It includes several sources of TeV $\gamma$-rays, seen by Milagro \cite{mgro}, which could also 
radiate neutrinos. These are potentially visible from IceCUBE location and two of them deserve a special mention:
\begin{itemize}
\item
{MGRO 2019+37} is still unidentified and there is no visible 
correlation with matter excess. ARGO \& Veritas do not see it.
Fitting 
\begin{equation}
I_\gamma=10^{-11}\times E^{-2.2}\times e^{-\sqrt{E/E_c}}/\mbox{TeV cm$^2$ s}
\end{equation}
with $E_c=45$ TeV, it could imply up to
1.5 muon events per km$^2$ year above 1 TeV. 
\item
{MGRO 1908+06} is seen also by ARGO$\approx$Milagro$>$HESS; a pulsar 
has been found by Fermi-GLAST there. Using 
\begin{equation}
I_\gamma=2\times 10^{-11}\times E^{-2.3}\times e^{-\sqrt{E/E_c}}/\mbox{TeV cm$^2$ s}
\end{equation} 
with $E_c=30$ TeV, 
it could imply up to 
2.5 muon events  per km$^2$ year above 1 TeV.
\end{itemize}
A few remarks:
1) The cutoffs $E_c$ have been introduced to account for the 
CASA-MIA bounds at 100 TeV.
2) There is another source, MGRO 2032+41, but weaker. 
3) Let us repeat that these objects are weaker theoretical cases than SNR's, 
but at least, they include a large target material.

\noindent  {\bf Fermi Bubbles:}
Possible  cases of  diffuse/wide sources could be given by the recently discovered 
`Fermi bubbles'. Are they  a reservoire of 
galactic cosmic rays?  If so, they could be also
promising neutrino sources \cite{croka}. 
By extrapolating the flux to 
\begin{equation}
I_\gamma(E)\!=\!\Omega\; 10^{-9}
e^{-\sqrt{E/E_c}}/E^{2}/\mbox{ TeV cm$^2$ s}
\end{equation}
with $E_c=$100 TeV (meaning a cut at 1 PeV in CR spectrum) and using an angular size of 
$\Omega=0.2 \mbox{ \rm sr}\approx \pi\times (15^\circ)^2$,
we get a  signal of about 100 muons a year for 1 km$^2$ detector area. This could be observable in Km3NET as a wide neutrino source over the background.

\section{Summary and outlook}

The promising galactic neutrino sources are intimately tied to $\gamma$'s 
above 10 TeV.
The main points regarding the predictions  
of  the high-energy neutrino fluxes from galactic 
sources  are simply two:
\begin{itemize}
\item There are only few bright $\gamma$-ray sources, 
that correspond to detectable neutrino sources.
\item For these sources,  the 
expected neutrino signal is modest, even  assuming that the $\gamma$-ray 
emission is fully contributed by hadronic processes.
\end{itemize}
The studies of $\gamma$-rays from RX J1713.7-3946 prove
that it is possible to proceed toward reliable expectations for high-energy $\nu$s.
The future years will offer us good occasions of progress in this sense:
\begin{enumerate}
\item Multiwavelength observations and theory 
will help us to understand better the young SNR's and to 
quantify their hadronic and leptonic emissions.
\item
Sub-degree pointing with very high energy $\gamma$-rays 
could permit us to further test the paradigm of SNR+MC 
association for RX J1713.7-3946.
\end{enumerate}
These considerations emphasize 
the relevance of certain lines of progress, and suggest to 
increase synergies and collaboration 
between neutrino and gamma ray search.



\appendix

\section{Neutrino oscillations}

\paragraph*{Averaged oscillation probabilities:}
Three flavor oscillations are well-understood and {\em relevant}. 
For the transparent sources we are discussing, 
the simplest regime \cite{gp} applies and the relevant probabilities are 
just constant:
\begin{equation}
P_{\ell \ell'}=\sum_{i=1}^3 |U_{\ell i}^2|  |U_{\ell' i}^2| \ \ \ \ell,\ell'=e,\mu,\tau
\end{equation}
In order to provide explicit expressions of the averaged probabilities,
we adopt the standard parameterization of the leptonic 
mixing matrix to solve the unitarity contraints. Then, defining  
\begin{eqnarray}
\epsilon&=&\sin^2\theta_{13}\\ 
\alpha&=&\sin\theta_{13}\cos\delta \sin 2 \theta_{12} \sin 2 \theta_{23}\\ 
\beta&=& (1-\epsilon) \cos 2\theta_{12}/2 + \gamma \\
\gamma&=&(1+\epsilon) \cos 2 \theta_{12} \cos 2 \theta_{23}/2 
\end{eqnarray}
we obtain relatively simple expressions for the averaged probabilities:
\begin{eqnarray}
P_{ee}&=&(1-\epsilon)^2\left(1-\frac{1}{2} \sin^2 2\theta_{12} \right) +\epsilon^2 \\
P_{e\mu}&=&\frac{1-\epsilon}{2} \left[\cos^2 \theta_{23}  \sin^2 2\theta_{12} + 
4 \epsilon \sin^2 \theta_{23} \times \right. \nonumber \\
& \times& \left. \left(1-\frac{\sin^2 2\theta_{12}}{4}\right)   +  \alpha \cos 2\theta_{12} \right]  \\ 
P_{\mu\mu}&=&\frac{1}{3}  + \frac{3}{2}  \left[(1-\epsilon) \sin^2\theta_{23}-\frac{1}{3}
 \right]^2  + 
 \frac{1}{2} (\alpha- \beta)^2 \\
 P_{\mu\tau}&=&\frac{1}{8}\left[ (1+\epsilon)^2 +(1-\epsilon)^2(3  \sin^2 2 \theta_{23} - \sin^2 2 \theta_{12} ) \right]+\nonumber\\
&-& \frac{1}{2} (\alpha-\gamma)^2
\end{eqnarray}
One can check explicitly that the diagonal probabilities obey $1\ge P_{\ell\ell} \ge 1/3$ and 
the out-of-diagonal ones $1/2\ge P_{\ell\ell'} \ge 0$.
The expressions of the 
 remaining probabilities  
 $P_{e\tau}$ and $P_{\tau\tau}$ are obtained from these 
 expressions,
 mapping $\delta\to \delta +\pi$ and $\theta_{23}\to \pi/2-\theta_{23}$.
Note that these formulae 
depend on $\epsilon$, on 
the `doubled angles' $2\theta_{12}$ and 
$2\theta_{23}$,  and on the combination $\sin\theta_{13} \cos\delta$.

\begin{figure}[t]
\centerline{\includegraphics[width=5cm]{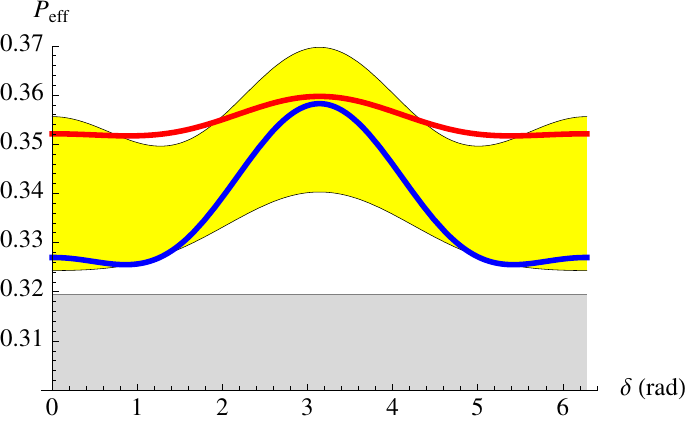}}
\caption{\em\footnotesize Dependence of the effective oscillation probability on the CP-violating phase and other oscillation parameters; 
see the text for the meaning of the various lines. \label{figa}}
 \end{figure}

The muon neutrino/antineutrino flux
after oscillation $F_\mu$, as a function of the fluxes before oscillation,
$F^0_{e,\mu}$, is:
\begin{equation}
F_\mu=P_{\mu\mu} F_\mu^0 +P_{e\mu} F_e^0\equiv  P_{\mbox{\tiny eff}}\ F_\mu^0 \label{epo}
\end{equation}
Two important remarks on $P_{\mbox{\tiny eff}}$ are in order:\newline
1) $P_{\mbox{\tiny eff}}$ depends on the energy through the flux ratio $F_e^0/F_\mu^0$.
A typical value for CR  collisions is $1/2$, 
for when we consider an almost equal amount of charged pions $\pi^\pm$, 
1 electron (anti)neutrino each 2 muon (anti)neutrinos are 
produced, all with similar energies (in \cite{fv08} one 
finds more precise statements for power law distributions).
However, other cases are possible at least at a speculative level: e.g., 
if muons are absorbed before decaying (which could happen, e.g., in microquasars) 
the initial beam is purely made of muon neutrinos,
$F_e^0/F_\mu^0=0$; or, if the neutrino originate from neutron decays, 
only electron antineutrinos are produced
$F_e^0/F_\mu^0=\infty$. \newline
2) The probability $P_{\mbox{\tiny eff}}$ 
depends also, to some extent, on the 
 most poorly known oscillation-parameters. Fig.~\ref{figa} 
 shows the variation of $P_{\mbox{\tiny eff}}$ with the 
CP-violating phase $\delta$, where $\theta_{12}=34^\circ$ is well-known and thus fixed, 
whereas the other angles are moved in the wide and conservative  
ranges $\theta_{23}=(40^\circ\to 50^\circ)$ and $\theta_{13}=(6^\circ\to 12^\circ)$--wiggly yellow band--and we set
$F_e^0/F_\mu^0=1/2$. For comparison, we show also the curves obtained by setting 
$\theta_{23}=50^\circ$ and $\theta_{13}=6^\circ$ (upper, red) and 
$\theta_{23}=40^\circ$ and $\theta_{13}=12^\circ$ (lower, blue); the lower part, 
grey region, is forbidden. 
$\theta_{13}$ is a relevant variable, thus future precise measurements will 
impact somewhat on oscillation probabilities.

\paragraph*{Numerical values:}

The values of $P_{\ell\ell'}$ change a bit with
the new best fit values of $\theta_{23}=40^\circ$ and   
$\theta_{13}=9^\circ$ of \cite{lisi} (assuming the new reactor fluxes) 
along with \cite{thomas} $\delta=-90^\circ$, that replace the previous reference (or extremal) 
values $\theta_{23}=45^\circ$ and $\theta_{13}=0^\circ$, now disfavored;
$\theta_{12}$ changed slightly $35^\circ\to 34^\circ$. 
The new values imply that the symmetric matrix with 
elements $P_{\ell\ell'}$ changes as follows;
\begin{equation}
\left(
\begin{array}{ccc}
0.558 & 0.221 & 0.221 \\
& 0.390 & 0.390 \\
&& 0.390 
\end{array}
\right)
\to
\left(
\begin{array}{ccc}
0.543 & 0.262 & 0.195 \\
& 0.364 & 0.374 \\
&& 0.430
\end{array}
\right)
\end{equation}
The changes are not very large, though.

The present ranges of parameters, approximatively 95\%, 
are 
$\theta_{12}=32^\circ-36^\circ$,
$\theta_{13}=4^\circ-12^\circ$ and
$\theta_{23}=37^\circ-51^\circ$, compare \cite{lisi} and \cite{thomas}. 
The precise values of the best fit and of the ranges depend on the 
details of the analysis.\footnote{In particular, there is an important dependence on the (theoretical and experimental) value of the flux from the reactors, discussed in \cite{thomas}, that is another reason why
the measurements of Double CHOOZ, Reno and DAYA Bay will be important to proceed further.} 
For illustration, we adopt in the following the comparably larger 
value of $\theta_{13}$ and smaller central value of $\theta_{23}$ of \cite{lisi},  
and take into account also 
the existing (feeble) indication for $\delta<0$ \cite{lisi,thomas},
assuming:
\begin{equation}
\begin{array}{cc}
\theta_{12}=34^\circ \pm 1^\circ, &   
\sin^2\theta_{13}=0.025 \pm 0.007,\\\  
\theta_{23}=42^\circ\pm 3.5^\circ, &
\delta=-\frac{\pi}{2} \pm \frac{3 \pi}{4} \label{prs}
\end{array}
\end{equation}
that allow us to perform simple statistical analyses. More precisely, we can
consider these  
parameters as independent Gaussian variables, 
and constrain the averaged probabilities as described in \cite{vulcano}.

\begin{figure}[t]
\centerline{\includegraphics[width=4.2cm,angle=0]{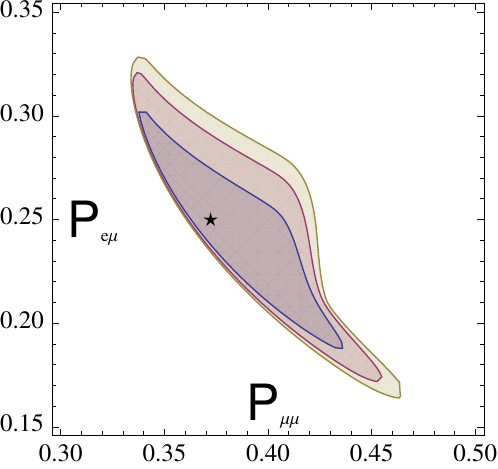}}
\caption{\em\footnotesize 
Values of $P_{\mu\mu}$ and $P_{e\mu}$, compatible at 
$1\sigma$, 90\% and $2\sigma\approx 95\%$ with the present 
information on the mixing angles, as described in Eq.~\ref{prs}.
\label{fig15}}
\end{figure}

Now, let us consider the allowed ranges  for the two most 
important averaged probabilities, $P_{e\mu}$ and $P_{\mu\mu}$. From Eq.~\ref{prs}, we 
 derive at 1$\sigma$ the values $P_{\mu\mu}=0.37^{+0.05}_{-0.03}$ and 
$P_{e\mu}=0.25^{+0.04}_{-0.05}$.
Fig.~\ref{fig15} shows that 
these two quantities are strongly correlated; thus, it is not immediate 
to derive the expected range of $P_{\mbox{\tiny eff}}$. 
This can be done however propagating the error directly; 
the result is shown
in Tab.~\ref{tabq}. For the important case $F_e^0/F_\mu^0=1/2$, we find the 2$\sigma$ range
$P_{\mbox{\tiny eff}}=0.32-0.36$, which is wider but 
does not disagree severely with the result in 
\cite{fv08}, $P_{\mbox{\tiny eff}}=0.33-0.35$.

All in all, the new results  may eventually lead to a small increase of  
$P_{\mbox{\tiny eff}}$, and thus of the signal: about  $5\%$. In view of the larger particle physics uncertainties, and much larger astrophysical ones, we maintain our position, expressed in 2004 \cite{fv04}, 
that the small uncertainties in the oscillation parameters imply that
``there is little hope to learn anything useful on 3 flavor oscillations'' 
studying high energy neutrinos from cosmic sources. 
Different opinions have been however expressed several times  in the 
literature, and authoritatively, in \cite{ks}; our criticisms to this 
specific proposal can be found in \cite{vulcano}, footnote 1.

 \paragraph*{Expansion in small parameters:}
It is also possible to expand the averaged probabilities $P_{\ell \ell'}$
in $\sin\theta_{13}$ and 
$\cos(2\theta_{23})$ that are small parameters 
and, due to the present uncertainties, vary in a similar range; in fact,  
$\sin( 10^\circ)=\cos(2\times 40^\circ)=-\cos(2\times 50^\circ)\sim 1/6$.
Thus, we expand  the averaged probabilities 
around the point $\theta_{13}=0$ and $\theta_{23}=45^\circ$
(namely, where the two small parameters are zero)
neglecting cubic terms and 
finding:
\begin{eqnarray}
P_{ee} &=& 1-\frac{x}{2}-2 w \ , \ P_{\mu\tau} = \frac{1}{2} -\frac{x}{8} - z  \nonumber \\ 
P_{e\mu} &=& \frac{x}{4}+y+w\ ,\ P_{\mu\mu} = \frac{1}{2} -\frac{x}{8}-y -w +z\\
P_{e\tau} &=& \frac{x}{4}-y+w  \ , \ P_{\tau\tau} = \frac{1}{2}  -\frac{x}{8} +y - w +z  \nonumber
\end{eqnarray}
We introduced for convenience various quantities of different size,
\begin{eqnarray}
\mbox{\small order 1: } x  &\equiv&  \sin^2 (2\theta_{12})   \\
\mbox{\small linear: } y  &\equiv&  \frac{1}{4} 
\left[ \sin \theta_{13} \cos\delta \sqrt{x(1\!-\! x)} \!+\!\cos(2\theta_{23}) x \right]  \\
\mbox{\small quadratic: } w  &\equiv&  \sin^2\! \theta_{13}\ (1\! -\!  x/2)   \\
\mbox{\small quadratic: } z  &\equiv& \frac{1}{2}  \left[\sin^2\! \theta_{13}\ (1\!+\! \cos(2\delta) x/2)  +
\cos^2 (2 \theta_{23})\times \right. \nonumber \\ 
   (1\! -\! x/4) &-& \left. \sin \theta_{13} \cos (2 \theta_{23}) \cos\delta \sqrt{x(1\!-\! x)}
  \right] 
\end{eqnarray}
the linear expansion has been obtained already in \cite{fv04}.
Several remarks are in order:
\begin{enumerate}
\item By summing $x,y,w$ or $z$ in any row or column of the symmetric matrix
with elements $P_{\ell\ell'}$ one gets zero; 
this is a consequence of unitarity.
\item The unique linear term $y$ involves a weighted sum 
of $\sin\theta_{13}$ and $\cos(2\theta_{23})$; this implies that there is a strong 
correlation between the linear variation of $P_{e\mu}$ and $P_{\mu\mu}$, while 
$P_{ee}$ and $P_{\mu\tau}$ do not vary at this order. Note that due to the coefficients, 
the impact of $\cos(2\theta_{23})$ is larger than the one of $\sin\theta_{13}$.
\item The relative size of $y$ and $z$ depends on $\delta$; the latter can be larger than the former (when $\delta\sim \pi$) even if these two terms are formally of different orders.
\end{enumerate}
These expressions can be useful for some purposes, e.g., to understand the correlations among the 
probabilities, however they are not much simpler than the 
full expressions. Moreover, the a priori limits $1\ge P_{\ell\ell} \ge 1/3$ and 
$1/2\ge P_{\ell\ell'} \ge 0$ (when $\ell\neq \ell'$) and the specific values of the parameters of oscillations imply that $P_{\mbox{\tiny\rm eff}}$ is close to the physical boundary \cite{fv08}; in other terms, 
the non-linear terms have a certain role, as consistent with the previous last considerations. 
For these reasons, we decided to use the full expressions of $P_{\ell\ell'}$ for the above 
analysis. 

\begin{table}[t]
\centerline{
\begin{tabular}{|c||c|c|c|c|c|} 
\hline
$F_e^0/F_\mu^0$ & 0 & \bf 1/2 & 1 & 2 & $\infty$ \\ \hline
$P_{\mbox{\tiny eff}}$ & $0.37^{+0.05}_{-0.03}$ & $\bf 0.33^{+0.02}_{-0.01}$  & $0.31^{+0.01}_{-0.00}$  & $0.29^{+0.02}_{-0.01}$  & $0.25^{+0.05}_{-0.05}$  \\ \hline
$\delta P_{\mbox{\tiny eff}}/P_{\mbox{\tiny eff}}$ & 11\% & \bf 4\% & 3\% & 6\% & 18\%\\ \hline
\end{tabular}}
\caption{\em\footnotesize Value of the effective probability of oscillation, Eq.~\ref{epo}, 
and of its error, for various values of the flavor ratio $F_e^0/F_\mu^0$.\label{tabq}}
\end{table}

\end{document}